\documentstyle [prl,preprint,aps] {revtex}

\begin{document}
\draft %revtex
\preprint{Applied Physics Report No. 93-21}
         %$\bigsqcup\bigsqcup$}
\title{
  Persistent Current of a One-Dimensional Wigner Crystal-Ring
}
\author{I. V. Krive,$^{(1,2)}$ R. I. Shekhter,$^{(1)}$
S. M. Girvin,$^{(3)}$ and M. Jonson${^(1)}$}
\address{
     $^{(1)}$Department of Applied Physics,
     Chalmers University of Technology and G{\"o}teborg University,
     S-412 96 G\"oteborg, Sweden    \\
$^{(2)}$B. Verkin Institute for Low
Temperature Physics and Engineering, Academy of Sciences of Ukraine,
310164 Kharkov, Ukraine \\
$^{(3(}$Department of Physics, Indiana University, Bloomington, Indiana
47405, USA }
\maketitle
\begin{abstract}
We calculate the magnetic moment (`persistent current') in a strongly
correlated electron system --- a Wigner crystal --- in a one-dimensional
ballistic ring. The flux- and
temperature dependence of the persistent current is
shown to be essentially the same
as for a system of non-interacting electrons.
In contrast, by incorporating into the ring geometry
a tunnel barrier, that pins the Wigner
crystal, the current is suppressed and its
temperature dependence is drastically changed. The competition between
two temperature effects --- a reduced barrier height for macroscopic tunneling
and a loss of quantum coherence --- results in a
sharp peak in the temperature dependence, which for
a rigid Wigner crystal appears at
$T\sim 0.5\ \hbar s/L$, ($s$ is the sound velocity of the Wigner crystal,
$L$ is the length of the ring).
\end{abstract}
%\maketitle
\narrowtext\noindent
In recent experiments \cite{one,two} persistent currents have been
observed in the ballistic transport regime of mesoscopic
rings formed in the laterally
confined two-dimensional electron gas of certain AlGaAs heterostructures.
The current $I$ and the associated magnetic moment were found to oscillate
as a function of magnetic flux with period
 $\Phi_0=hc/e$ --- the quantum unit of flux ---  and amplitude
$I_0\sim ev_F/L$ ($e$ is the electronic charge, $v_F$ the Fermi velocity,
and $L$ the length of the circumference of the ring). These results are in
excellent agreement with a theory of such Aharonov-Bohm (AB)
oscillations based on a free electron model of the ballistic electrons
\cite{three,four}. Since electron-electron interactions in the
semiconductor ring are not weak, and since electron correlations must play
an important role when the density of conducting electrons is low, this
agreement is quite surprising. In diffusive metal rings, for example, where
the electronic mean free path is short ($\ell\ll L$), it has been
suggested that electron correlations significantly enhance the
amplitude of the AB oscillations \cite{five}. Thus the question of how
Coulomb correlations in a system of ballistic electrons affect the
magnitude of the persistent current is of significant interest.

In this Letter we study persistent currents and AB oscillations in
systems of spinless interacting electrons confined to a one-dimensional
ring; the
electrons are assumed to be so strongly correlated that they form
a Wigner crystal. In an ideal ring the mechanism of the persistent
current is a dissipationless sliding of the crystal as a whole. We
demonstrate that the resulting current oscillates as a function of
magnetic flux with period $\Phi_0$. Its amplitude at low
temperatures is exactly the same, $I_0=ev_F/L\; (v_F = \pi\hbar/ma$ is of the
Fermi velocity, $a$ is the period of the Wigner
crystal, $m$ is the electron mass) as for noninteracting electrons
of the same density in accordance with general theorem \cite{MullerG}. If the
temperature is raised, the
amplitude of the oscillations is exponentially suppresssed:
 $I(T) \sim I_0 \exp(-\pi T/2 T_0)$, where $T_0 \equiv \hbar v_F/L$ is the
characteristic crossover temperature. Therefore the
magnitude as well as the temperature dependence of the persistent current
carried by an ideal Wigner crystal  looks completely
identical to that of a current carried by a free electron gas.

The situation changes drastically if a potential barrier, somewhere along
the ring, impedes the motion of the electrons. Charge transport in this
case requires that electrons tunnel through the barrier - the process
which for strongly correlated electrons can be viewed as a macroscopic
tunneling of a Wigner crystal-ring. In the case of high enough barrier
(strong pinning) it is convenient
to think of the motion of the crystal as a two-step process, where first a
single electron tunnels through the barrier producing a deformation of a
finite portion of the Wigner crystal, which then is
relaxed \cite{six,seven}. This process
necessarily depends on the elastic properties of the crystal, and as
a result the magnitude of the persistent current will depend on the sound
velocity, $s$, in the Wigner lattice.
As our analysis below will show, the temperature
dependence of the amplitude of the AB oscillations is also affected in a
qualitative way.
The presence of the tunnel barrier,
which pins the Wigner crystal and makes charge transfer possible only by
macroscopic tunneling,
strongly decreases the zero temperature value of the persistent current since
for a repulsive interaction quantum fluctuations in a strongly correlated
electron system renormalize the barrier upward.
The finite ring circumference cuts off the divergent renormalization of the
barrier height which occurs in the thermodynamic limit of a Luttinger liquid
\cite{KF_PRB} or Wigner crystal \cite{six,seven}. Thus the persistent current
at zero temperature is greatly reduced but is still finite.
The competition between two effects of an increased temperature --- a
temperature stimulated  tunneling and a
loss of phase coherence due to the enhancement of destructive interference ---
leads to a sharp
maximum in the temperature dependence of the persistent current. For a rigid
crystal this maximum occurs at
$T\sim 0.5\ T_s$, where $T_s\equiv\hbar s/L\gg T_0$.
This effect makes it possible to measure the Wigner crystal sound velocity
in a ring with an `adjustable
barrier' (height controlled by a gate voltage).

The starting point of our analysis is a model system, where the Wigner
crystal is regarded as an elastic chain of spinless electrons forming a
ring. In the presence of a
potential barrier, smooth on the scale of $a$ but well localized on the scale
of $L$, the Lagrangian of such a
system in the long wavelength approximation  is \cite{seven}
\begin{equation}
L = \frac{ma}{8\pi^2}\left\{\dot \varphi^2 -
s^2(\varphi^\prime)^2\right\} - V_0\delta(x)\cos(\varphi) .
\label{lagrangian}
\end{equation}
Here $\varphi=2\pi u(x)/a$ is the dynamical displacement field of the
crystal and $V_0$ is the magnitude of the pinning potential (without loss of
generality placed at the point $x=0$).

We emphasize that (\ref{lagrangian}) is an
effective  Lagrangian  that describes the long wavelength aspects of
the quantum dynamics of the Wigner crystal.  The
short wavelength fluctuations do not affect the global dynamics of the
system, and only result in a renormalization of the magnitude $V_0$ of the
potential, (already included in (\ref{lagrangian}) but negligible for a
stiff
 Wigner crystal \cite{seven}).
We assume that the ring circumference is large enough to justify dropping
terms from (\ref{lagrangian}) which are irrelevant in an infinite system.

In the presence of a magnetic field, directed normal to the plane of the
ring, the one-dimensional Lagrangian (\ref{lagrangian}) acquires an
additional term, $L_{int}$. This term describes the AB interaction of the
Wigner crystal with the vector potential of an electromagnetic field,
$A_\varphi =\Phi/L$ ($\Phi$ is the magnetic flux through the ring). The AB
interaction term, rewritten using the real scalar displacement field
$\varphi$, has the form of a total time derivative,
\begin{equation}
L_{int} = \left(\hbar/L\right)\left(\Phi/\Phi_0\right) \dot\varphi,
\label{lint}
\end{equation}
and affects, as must be the case, only the quantum dynamics of the crystal.
%
% Equation~(\ref{lint}) has a term that depends on the parity $P_N\equiv
%[1-(-1)^N]/2$ of the number $N$ of electrons (can be odd or even).

The flux-induced persistent current $I(\Phi) = -c \partial F/\partial \Phi$,
is defined in terms of the sensitivity of the free energy of the ring to
a magnetic flux.
For the following analysis, it is convenient to express the free energy $F$
as a functional integral over quantum- and thermal fluctuations of
the displacement field,
\begin{equation}
%F = - T \ln \left\{\sum_{n=-\infty}^\infty \left(-1\right)^{n(N-1)}
%\int D\varphi_n \exp\left(-S_E[\varphi_n]/\hbar\right)\right\} ,
F = - k_B T \ln \left\{\sum_{n=-\infty}^\infty \left(-1\right)^{n(N-1)}
\int D\varphi_n {\rm e}^{-S_E[\varphi_n]/\hbar}\right\} ,
\label{F}
\end{equation}
where the action $S_E$ derives from the Lagrangian
(\ref{lagrangian}), (\ref{lint}) in the imaginary time representation.
`Twisted' boundary conditions in imaginary time
are imposed on the field $\varphi$ (see, e.g. \cite{eleven}) :
\begin{equation}
\varphi_n(\tau + \beta, x) = \varphi_n(\tau,x) + 2\pi n .
\label{boundcond}
\end{equation}
Here $n= 0, \pm 1, \pm 2 \ldots$ is the topological (winding) number,
classifying homotopically inequivalent trajectories. The physical meaning
of the boundary condition (\ref{boundcond}) follows from the definition
of the field $\varphi = 2\pi u(x)/a$; a uniform shift of the crystal by a
distance equal to an integer times the lattice constant $a$ leads, in the
ring geometry, to a
state identical to the initial state after
certain permutations of electrons. For the minimum shift by $1\times a$
$(\Delta\varphi = 2\pi)$,
the initial state is recovered after $(N-1)$ succesive permutations of pairs
of electrons.
The corresponding extra phase $\pi(N-1)$, that appears in the
many-particle  wavefunction because the electrons obey Fermi statistics,
generates the factor $(-1)^{n(N-1)}$ in (\ref{F}).
As we
will see below, this factor properly acounts for the parity effects in the
response of one-dimensional interacting electrons to a
magnetic field \cite{eight,nine,ten}.
We note in passing that the analogous twisted
boundary conditions appear when the Luttinger model is applied to a
ring geometry \cite{nine}. The appearance of the homotopic index $n$ in the
boundary condition (\ref{boundcond}) suggests that the functional integral
should first be calculated for trajectories belonging to a definite
homotopic class, and then the homotopically non-equivalent
classes of trajectories should be summed over.

In every homotopic class we will calculate the functional integral using the
saddle point approximation, assuming the saddle point action to be large,
$S_n\gg\hbar$, on the extremal trajectory given by the solution of the
classical equations of motion in imaginary time. Below we will show that
this assumption is justified for a stiff Wigner crystal $(\alpha\ll 1)$.

First we calculate the persistent current in the ideal, unpinned
crystal $(V_0=0)$. In a perfect (or weakly pinned) Wigner crystal,
long-wave quantum fluctuations are cut off at the wavelength of the order
of the crystal size $L$. It is physically evident that
we can imagine an ordered  crystal structure as long as the mean square
fluctations  of the
dimensionless field  $\varphi$,
\begin{equation}
		\langle \varphi^2 \rangle \sim \alpha
\int_{\pi/L}^{\pi/a}\frac{dk}{k}\coth\left(\frac{s\hbar}{2k_B T} k\right) ,
\label{five}
\end{equation}
are small so that $\langle \varphi^2 \rangle\ll 1$ ($T$ is the temperature,
$\alpha =\pi\hbar/msa$ is a dimensionless parameter that characterizes
the strength of quantum fluctuations in the Wigner crystal). For $T \to 0$
this restriction imposes an upper bound on the chain length $L\ll a{\rm
e}^{1/\alpha}$; for such  samples the thermal fluctations
are suppressed up to a temperature $T \alt T_s/\alpha$   ($T_s \equiv \hbar
s/L$). The situation is changed drastically for a strongly pinned Wigner
crystal
 where an
``intermediate" cut off scale appears \cite{seven}.

One can readily calculate the persistent current of an ideal ring as the
problem  in the long wavelength limit is described by a quadratic Lagrangian.
The extremal trajectory
corresponding to the boundary condition (\ref{boundcond}) is linear in
imaginary time and independent of the $x$-coordinate,
\begin{equation}
%\varphi_n(\tau) = \frac{2\pi n}{\beta} \tau
\varphi_n(\tau) = 2\pi n \left(\tau/\hbar\beta\right) .
\label{simplesol}
\end{equation}
By substituting (\ref{simplesol}) into
(\ref{lagrangian}, \ref{lint}, \ref{F}), it is easy to find an
exact solution
for the free energy in terms of the Jacobi function $\vartheta_3$  (see
e.g. \cite{twelve}). The asymptotic expressions
%\newpage\noindent
 for the persistent current
at high-
% $(T \gtrsim T_0)$
and low
%$(T \ll T_0)$
temperatures are
\begin{equation}
\frac{I_{WC}}{I_0} \simeq
\left\{
  \begin{array}{ll}
2\frac{T}{T_0}{\rm e}^{-\frac{\pi}{2}\frac{T}{T_0}}
    (-1)^N\sin\left(2\pi \frac{\Phi}{\Phi_0}\right), &
    T \gtrsim T_0  \\
    1-2\{\{\frac{\Phi}{\Phi_0} + \delta_N\}\}, & T\ll T_0
   \end{array}   \right.
\label{ifree}
\end{equation}
Here \{\{$x$\}\} denotes the fractional part of $x$, and the parity
dependent term $\delta_N$ is $1/2$ ($0$) for $N$ odd (even). Thus the
persistent current carried by an ideal Wigner crystal is a periodic function of
flux with period $\Phi_0 =hc/e$ and amplitude $I_0=ev_F/L$ at low temperatures.
The oscillations are exponentially damped at $T \gtrsim T_0=\hbar v_F/L$.
The current has a paramagnetic character when there
is an even number of electrons in the ring (i.e. the induced magnetic
moment is parallel to the external magnetic field) and diamagnetic for an
odd number of electrons. All these properties of the
persistent current coincide with those calculated using the model
of an ideal Fermi gas. For $T=0$, this was first shown in Ref.~\cite{MullerG}
for a general case of arbitrary Coulomb-like interaction.
At finite temperatures there are in general contributions due to
crystal deformations produced by thermally excited phonons.
It is possible to show \cite{fourteen} that, even in a perfect Wigner
crystal ring, the contribution of phonon
fluctuations to the action results in a correction to the persistent
current which is small if the temperature is less than $mv_0 s/2k_B$.

The thermal destruction of the persistent current can be characterized by a
crossover temperature $T_c$, where $ I\propto \exp(-T/T_{c}) $. From
(\ref{ifree}) one has  $T_{c}=(2/\pi)(\hbar v_{F}/L)$, which is twice as large
as the crossover temperature found for a ring of free electrons characterized
by
a constant chemical potential \cite{four,Cheung}. Rather than with the
electron-electron correlations ~\cite{nine} the factor of 2 difference is
connected  with the fact that in our case the number of electrons --- not the
chemical potential --- is fixed ~\cite{Grincwajg}.

Let us now consider the persistent current in a Wigner crystal in
the presence of
a potential barrier. A uniform sliding motion of the crystal is
impossible in this case, and charge transport along the ring
is connected with macroscopic quantum tunneling (MQT) of the Wigner
crystal. The character of the MQT is dictated by the pinning strength.
At strong pinning, $\alpha V_0 \gg T_s$, the mechanism for charge transport
around
the ring includes tunneling of a finite segment of the Wigner crystal through
the barrier, as well as the subsequent
 relaxation of the associated elastically deformed state of the crystal.
 In the weak pinning regime, $\alpha V_0 \ll T_s$, the Wigner crystal
as a whole tunnels
through the barrier (without essential distortions).
The above mechanisms for macroscopic tunneling were
first considered in connection with the tunneling of
commensurate charge density waves \cite{six} and have also been used
to describe the tunneling conductivity of a Wigner crystal \cite{seven}.
In these contexts it was shown \cite{six,seven} that in the case of
strong pinning the dominating tunneling process is the elastic
relaxation of the deformed state arising in the near-barrier region.

In the ring geometry a shift of the crystal as a whole by the lattice
period, $a$, may include one or several `tunneling steps' (by a
`tunneling step' we understand
the combined processes of tunneling and relaxation of the
elastic deformation).
In the case of strong pinning, the single-step tunneling is
described by the exact solution of the free equation of motion ($V_0=0$)
in imaginary time with the twisted boundary condition (\ref{boundcond}) and
$n=\pm 1$:
\begin{eqnarray}
\varphi_\sigma^{(s)}(\tau)&=& 2\sigma\arctan
\left[
  \coth\left(
  \frac{\pi|x|}{\hbar s\beta}
       \right)
\tan\left(\pi
\left(\frac{\tau}{\hbar\beta}-\frac{1}{2}\right) \right)
\right], \nonumber \\
\quad \sigma&=&\pm 1 .
\label{phisol}
\end{eqnarray}
A description of the dominating relaxation process in terms of this
`periodic instanton'-solution \cite{thirteen} is valid in the
region outside the interval $[-\ell_0, \ell_0]$ containing the part of the
crystal deformed by the inital tunneling process.
The length $\ell_0$, which is inversely proportional to the
potential $V_0$, appears only as a limit of the integration
over coordinate $x$; we assume that $\ell_0\ll L/2$, a criterion which one
can show to be equivalent to a restriction on temperature, $T\ll\alpha V_0$.

A multi-step solution is a sequence of single steps of type
(\ref{phisol}), corresponding to all possible intermediate rotations of the
crystal
\begin{equation}
\varphi_\sigma^{(m)}(\tau) = \sum_\ell
\varphi_{\sigma_\ell}^{(s)}(\tau-\tau_\ell), \quad
\sigma = \sum_\ell \sigma_\ell = \pm 1 .
\label{phim1}
\end{equation}
The set of solutions $\{\varphi_{\sigma}^{(m)}\}$,
corresponding to different configurations of single steps $\{\sigma_\ell\}$
and time-sequences
$\{\tau_\ell\}$ for the tunneling events, form the basis
in the well-known dilute instanton gas approximation.

At $T\ll T_s$, the multi-step solution (\ref{phim1}) can be used as the
extremal trajectory when calculating the partition function that
appears in the expression (\ref{F}) for the free energy.
In this manner we get the zero temperature value of the persistent current
as
\begin{equation}
I_{WC}(T=0) \sim (-1)^N
\frac{es}{L}\left(\frac{T_s}{\alpha V_0}\right)^{1/\alpha}
 \sin(2\pi \Phi/\Phi_0) .
\label{tzero}
\end{equation}
This result for the persistent current of a Wigner
crystal in the presence of a pinning potential barrier, clearly shows
that the effect of the barrier is simply to suppress the
zero temperature amplitude of the current. The net current depends on
the elastic properties of the Wigner lattice that reflects the fact
that for a strong pinning the charge transport in the ring is due  to
macroscopic quantum
 tunneling of the system through a deformed state of the crystal.

Except at very low temperatures, $T\ll T_s$, the only relevant saddle point
trajectory
is the single-step solution (\ref{phisol}).
By using it one  gets for the normalized current
\begin{eqnarray}
\frac{I_{WC}(T)}{I_{WC}(0)} &=& \frac{T}{T_s}\exp\left(\case{1}{\alpha}
f\left(T/T_s\right) \right), \quad  T \ll \alpha V_0
\nonumber \\
f(x) &=& \frac{\pi}{2} x - \ln\left( \frac{\sinh(\pi x)}{\pi x} \right) .
\label{thigh}
\end{eqnarray}
This result implies a non-monotonic temperature dependence  of the
persistent current; for
a stiff crystal (small $\alpha$) the current has an exponentially sharp
maximum at $T \sim 0.5 T_s$, with a width of the order of $\sqrt{T_0T_s}$.
The physical reason for this non-trivial temperature dependence ---
shown in Fig.~1 for different values of $\alpha$ ---
can be explained as follows:
It is easy to see from (\ref{phisol}) that
as the temperature is increased the picture of the elastic deformation
propagating as a `sharp' instanton changes (at $T\sim T_s$) into a picture of a
homogeneous sliding of the crystal as a whole.
This temperature-induced `softening' of the instanton reduces the
contribution to the action from the elastic deformation of the crystal.
Hence, the persistent current should increase with temperature. On the
other hand this effect competes with a thermal smearing of the phase
coherence  which --- as we showed for the unpinned crystal --- tends to
reduce the current. The sharp peak in the temperature
dependence of the persistent current carried by a strongly pinned Wigner
crystal,
is a result of this very competition.

At weak pinning a stiff Wigner crystal-ring tunnels through the barrier
as a whole and the persistent current does not depend on the elastic
properties of the chain. However, if the barrier is high ehough
$ \alpha T_{s} \alt V_{p} \ll T_{s}/\alpha $ (moderately weak pinning) the zero
temperature current is still greatly suppressed,
\begin{equation}
I_{wp}(T\rightarrow 0)\sim (-1)^N I_{0}\left(\frac{V_{0}}{T_{0}}
\right)^{1/4}\exp\left(-\sqrt{2\pi\frac{V_{0}}{T_{0}}}\right)
\sin\left(2\pi\frac{\Phi}{\Phi_{0}}\right) .
\label{iwp}
\end{equation}
This is in contrast to the case of non-interacting electrons where
even a large potential barrier (of the order of the Fermi energy) only leads
to a power-law supression of the persistent current ~\cite{Cheung}.

As temperature always makes tunneling easier, one may expect an
anomalous temperature behaviour of the persistent current even for a weakly
pinned Wigner crystal-ring. This is indeed the case, but unlike in the regime
of strong pinning, the maximum of the persistent current (which is now
attained at a potential-dependent temperature $ T^*\sim \sqrt{V_0T_0}$ )
is weakly pronounced. Therefore the only distinctive feature of the temperature
dependence of a moderately pinnned ($V_0 > T_0$) Wigner crystal-ring,
--- compared to that of a ring with noninteracting electrons --- is the shift
of
the crossover temperature to higher values, $ T_0\Rightarrow T^* > T_0$. For
very weak pinning, $V_p\ll T_0$, the response of a Wigner crystal-ring to a
magnetic flux is the same as for free electrons.

Formula (\ref{thigh}) is valid in the strong pinning limit, when
temperature is much smaller than $\alpha V_0$.
At high temperatures, $T \gtrsim \alpha V_0$, the pinning potential can
be treated as a perturbation when calculating the depinning of the Wigner
crystal. In this case we find unimportant
corrections to the persistent current in an ideal Wigner crystal (the
details of this calculation will be published elsewhere \cite{fourteen}).

By measuring the dependence of the persistent current on the
barrier height at zero temperature (\ref{tzero}) and its temperature
dependence (\ref{thigh}), one has an opportunity to
determine independently the stiffness parameter,
 $\alpha=h/2msa$, and
the sound velocity, $s$, in this system of strongly correlated electrons.
This gives us strong reasons to propose an
experiment using a gate-controlled barrier in a mesoscopic semiconductor
ring in order to study Wigner crystallisation and to measure the
parameters of the crystal.

%\newpage
In conclusion we have shown that in an ideal ring with no impurity
scattering, the persistent current carried by interacting electrons --- so
strongly correlated that they form a Wigner crystal --- is
indistinguishable from the current carried by a non-interacting Fermi gas.
By incorporating a potential barrier, $V_0 > \hbar v_F/L$,
in the ring structure, a qualitative change of the
magnitude and temperature dependence of the persistent current appears.
With an adjustable barrier, these differences can be used for
detecting and investigating the properties of the Wigner crystal.

% ------------------------- Acknowledgments
%%------------------------------------

We gratefully acknowledge discussions with L. Glazman,
A. Nersesyan, A. Sj{\"o}lander, and
A. Zagoskin.
This work was supported by the Swedish Royal Academy of Sciences,
the Swedish Natural Science Research Council, the Swedish
National Board for Industrial and Technical Development, by the
NSF through grant DMR-9113911, and by grant PH2-9187-0917 from International
Science Foundation.
One of us (I.K.) acknowledges the hospitality of the Department of Applied
Physics, CTH/GU.

% ------------------------- References
%%-----------------------------------------

\begin{figure}
\caption{Temperature dependence of the normalized persistent current in a
strongly pinned Wigner crystal of different stiffness (measured by
$\alpha^{-1}=2msa/h$; $T_s=\hbar s/k_BL$, see text). The sharp peak for stiff
crystals is a result of a
competition between two effects of temperature: a reduced renormalized
tunneling barrier and a loss of quantum coherence.
}
\end{figure}

\end{document}